\documentclass[prb,twocolumn,aps,showpacs]{revtex4}
\usepackage{graphics,graphicx,epsfig}
\begin{document}
\title{Anomalous roughening in competitive growth models with time-decreasing rates of correlated dynamics}
\author{F. D. A. Aar\~ao Reis}
\affiliation{
Instituto de F\'\i sica, Universidade Federal Fluminense,\\
Avenida Litor\^anea s/n, 24210-340 Niter\'oi RJ, Brazil}
\date{\today}

\begin{abstract}

Lattice growth models where uncorrelated random deposition competes with some aggregation dynamics
that generates correlations are studied with rates of the correlated component decreasing as a power law.
These models have anomalous roughening, with anomalous exponents related to the normal exponents of
the correlated dynamics, to an exponent characterizing the aggregation mechanism and to that
power law exponent. This is shown by a scaling approach extending the Family-Vicsek
relation previously derived for the models with time-independent rates, thus providing a connection of 
normal and anomalous growth models. Simulation results for several models support those conclusions.
Remarkable anomalous effects are observed even for slowly decreasing rates of the correlated component,
which may correspond to feasible temperature changes in systems with activated
dynamics. The scaling exponents of the correlated component can be obtained only from the estimates of
three anomalous exponents, without knowledge of the aggregation mechanism, and a possible application
is discussed. For some models, the corresponding Edwards-Wilkinson and
Kardar-Parisi-Zhang equations are also discussed.

\end{abstract}

\maketitle

\section{Introduction}
\label{intro}

The technological interest in thin films, multilayers and related nanostructures
motivated the study of continuous and atomistic growth models
and kinetic roughening theories \cite{mbe,barabasi,krug}. In the simplest
cases, surface fluctuations follow the standard
Family-Vicsek (FV) scaling relation \cite{barabasi,fv}. The amplitudes of fluctuation modes
saturate at times increasing with the wavelength and the global roughness saturates when the
largest excited mode is of the order of the substrate size. Local and global surface fluctuations
scale with the same exponents. This is called normal roughening.
On the other hand, a large number of systems show anomalous roughening
\cite{huo,saitou,hasan,schwarzacher,auger,fu,lafouresse},
where local slopes continuously increase and local exponents differ from the global ones
\cite{schwarzacher,lopez}. Theoretical approaches have shown various conditions, such as
symmetries and conservation laws, where anomalous scaling (AS) is expected \cite{lopezPRL2005,pradas}.

With the advance in deposition techniques, changing physico-chemical conditions during the growth
of a film is possible. In vapor methods, this can be achieved by varying pressure or temperature.
In electrodeposition, the evolution of surface morphology changes the electric field distribution and,
consequently, the local reaction rates may change. Thus, local growth is a competition between different
dynamic processes (adsorption, reaction, surface diffusion) with time-dependent rates. As a simple
illustration, one may consider a constant flux of atoms or molecules and time-varying rates of surface
processes (responsible for fluctuation correlations). These features may have significant effects on
roughness scaling. For instance, in Ref. \protect\cite{chou}, it was shown that a sudden
change in growth conditions of an Edwards-Wilkinson \cite{ew} model may lead
to power-law relaxation between different growth regimes. Moreover, Ref. \protect\cite{pradas} shows
that stochastic growth equation models with time-dependent couplings have AS under certain conditions.

Those findings motivate the present investigation of lattice models with competition of two aggregation
dynamics with time-dependent rates. We consider limited mobility models where uncorrelated
random deposition competes with some type of correlated component whose
rate decreases in time as a power-law.
These systems show AS, in contrast to the normal scaling observed
in the original models with constant rates \cite{rdcor}. 
The anomalous exponents are related to the (normal) exponents of the correlated growth dynamics and
to the exponent of the time-decreasing rate. Those relations are obtained by direct substitution of the
time-dependent rate in the FV relation of the model with time-independent rates, which provides
a connection between scaling relations of normal and anomalous models. Those conclusions are
supported by simulation
results. Remarkable anomalous effects are observed even for slowly decreasing rates,
which may correspond e. g. to small changes of temperature in systems with activated
dynamics. For some models, the general forms of the corresponding Edwards-Wilkinson (EW) \cite{ew} and
Kardar-Parisi-Zhang \cite{kpz} equations are also discussed.

The hydrodynamic equations associated to our growth models have time-dependent couplings, thus the
existence of AS is predicted by the approach of Ref. \protect\cite{pradas}.
However, the aim of  work is not discussing conditions for AS, but to explore details of a class of lattice
models with realistic dynamics showing that feature, with a simple connection of normal and
anomalous roughnening and an important role of the aggregation mechanism.
A possible application of our model is discussed, as well as systems where it is
certainly not applicable.

The rest of this work is organized as follows. In Sec. II, we review the scaling concepts for
competitive models with correlated and uncorrelated components and extend the approach
to the FV relation of the height-height correlation function. In Sec. III, we extend that
approach to competitive models with time-decreasing rates of the correlated component and calculate
their anomalous exponents. Sec. IV shows simulation results of some models in two-dimensional
substrates, confirming the theoretical predictions. In Sec. V, we discuss the EW equations
(in all dimensions) and KPZ equations (in $1+1$ dimensions) associated to some of those models.
In Sec. VI, we discuss the relation with experimental work and the reliability of our models.
Sec. VII summarizes our results and conclusions.

\section{Lattice models with competition of correlated and uncorrelated deposition}
\label{scalingcompetitive}

In these models, cubic particles of size $a$ are sequentially deposited on randomly chosen columns
of a $d$-dimensional substrate of lateral size $L$, in a direction perpendicular to the substrate.
The time interval for deposition of one layer of atoms [${\left( L/a\right)}^d$ atoms] is $\tau$.
Each incident particle has two possibilities: 
(1) permanently stick at the top of the column of incidence, with probability $1-p$
(uncorrelated deposition - UD);
(2) execute some movement (diffusion, desorption, or eventually stick at some point of the
column of incidence) according to an aggregation rule that takes into account the neighboring
column heights, with probability $p$. This second component
generates correlations of the column heights and is hereafter called correlated deposition (CD).

\begin{figure}
\includegraphics[width=7cm]{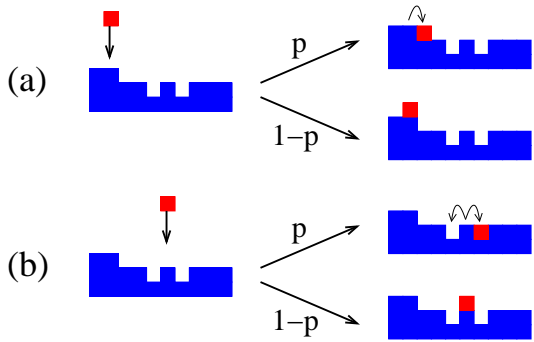}
\caption{(Color online) Illustration of the model with competition between
RDSR (probability $p$) and UD (probability $1-p$) in one-dimensional substrates.
In (a), after the choice of RDSR, the incident particle attaches at the column
at the right side. In (b), after the choice of the RDSR,
the particle randomly chooses between the neighboring columns (curved arrows);
only the attachment at column at the right side is illustrated. In (a) and (b),
after the choice of UD, the particle sticks at the top of the column of incidence.}
\label{fig1}
\end{figure}

Fig. 1 illustrates the model where the correlated component is
random deposition with surface relaxation (RDSR) \cite{family}.
In RDSR, the particle sticks at the top of the column of incidence if no neighboring
column has a smaller height, otherwise it sticks at the top of the column with the
smallest height among the neighbors (in case of a draw, one of the columns with
the smallest height is randomly chosen).
The competitive model UD-RDSR and another one involving UD and ballistic deposition (BD) \cite{vold}
were introduced by Albano and co-workers
\cite{albano1,albano2}.

In this section, we consider the case where the probability $p$ is kept constant
during the film growth.

The global roughness of the film surface is defined as the
rms fluctuation of the height $h$ around its average position $\overline{h}$:
\begin{equation}
W(L,t)\equiv {\left[ { \left<  {\left( h - \overline{h}\right) }^2  \right> }
\right] }^{1/2} ,
\label{defw}
\end{equation}
where the overbars indicate spatial averages and the angular brackets indicate
configurational averages.
At short times, UD dominates, thus the roughness increases as
\begin{equation}
W_{RD}\approx a{\left( t/\tau\right)}^{1/2} .
\label{wrandom}
\end{equation}
After a crossover time $t_c$, the CD determines the universality class of the process,
characterized by a set of exponents hereafter indexed with $C$.
The roughness follows Family-Vicsek (FV) scaling \cite{fv} as
\begin{equation}
W(L,t) \approx A L^{\alpha_C} f\left( \frac{t}{t_\times}\right) ,
\label{fv}
\end{equation}
where $\alpha_C$ is the roughness exponent, $f$
is a scaling function such that $f\sim 1$ in the regime of roughness saturation
($t\to\infty$) and $t_\times$ is the characteristic time of crossover to
saturation, which scales as 
\begin{equation}
t_\times \approx BL^{z_C} , 
\label{scalingttimes}
\end{equation}
where $z_C$ is the dynamic exponent. For $t\ll t_\times$ (but $t\gg t_c$),
the roughness scales as
\begin{equation}
W\approx Ct^{\beta_C} , 
\label{scalingwgr}
\end{equation}
where $\beta_C=\alpha_C/z_C$ is the growth exponent. In this
growth regime, $f(x)\sim x^{\beta_C}$ in Eq. (\ref{fv}).

The exponents $\alpha_C$, $\beta_C$ and $z_C$ depend on the basic symmetries of the
CD, but the amplitudes $A$, $B$ and $C$ are model-dependent.
For small $p$, they scale as
\begin{equation}
A\sim p^{-\delta} ,
\label{defdelta}
\end{equation}
\begin{equation}
B\sim p^{-y} ,
\label{defy}
\end{equation}
and 
\begin{equation}
C\sim p^{-\gamma} ,
\label{defgamma}
\end{equation}
where the convention of crossover exponents $\left( \delta,y,\gamma\right)$ of Albano and co-workers
\cite{albano1} was used.
FV scaling implies \cite{albano1}
\begin{equation}
y\beta_C -\delta+\gamma = 0 .
\label{scalingalbano}
\end{equation}

Different scaling approaches  have already explained the values of the exponents obtained in
simulations of lattice models \cite{rdcor,albanordcor}.
Below we review the scaling approach of Ref. \protect\cite{rdcor} for the global roughness
and extend it to the height-height correlation function (HHCF) and the local roughness.

For $p\ll 1$, most deposited atoms attach to the top of the randomly chosen column (UD).
Thus, the height difference of neighboring columns, $\Delta h$, is of order
$\Delta h \sim a{\left( \Delta t / \tau\right)}^{1/2}$ after a time interval $\Delta t$
(as Eq. \ref{wrandom}). On the other hand, the average time for a CD event (probability $p$)
to take place at a given column is $\tau_c\sim\tau/p$.

If CD is BD, that event immediately creates correlations between the neighboring columns.
Thus, the time of crossover from UD to correlated growth is
\begin{equation}
t_c \sim \tau_c \sim p^{-1}\tau \qquad (BD-like) .
\label{scalingtc1}
\end{equation}
This applies to other BD-like models, as discussed e. g. in Ref. \protect\cite{bbdflavio}.

However, if CD is a solid-on-solid (SOS) model, such as RDSR, a single correlated event does not
cancel the random fluctuation of neighboring column heights $\Delta h$. Instead, that fluctuation is
cancelled only when the number of correlated events $N_c$ is of order ${\Delta h}/a$.
At the crossover time $t_c$, that number is $N_c = t_c/\tau_c \sim t_c p/\tau$. Thus
\begin{equation}
t_c \sim p^{-2}\tau \qquad (SOS)
\label{scalingtc2}
\end{equation}

In both cases, all time scales of the purely correlated system ($p=1$) are also changed by the scaling
factor $t_c/\tau$, such as the saturation time $t_\times$ (Eq. \ref{scalingttimes}). Thus
we have $y=1$ for BD-like models and $y=2$ for SOS models in Eq. (\ref{defy}), independently of
the class of the CD and substrate dimension (this improves the picture that emerged from 
a former work \cite{lam}, which restricted $y=1$ to KPZ and $y=2$ to EW).

The average height difference between neighboring columns saturates at
$\Delta h_c \sim a{\left( t_c / \tau\right)}^{1/2}$, which is of order $ap^{-y/2}$ both for BD-like
and SOS models. This is the scaling factor for global height fluctuations (Eqs. \ref{fv} and \ref{defdelta}), 
consequently
\begin{equation}
\delta =y/2 .
\label{deltay}
\end{equation}
It gives $\delta =1/2$ for BD-like models and $\delta =1$ for SOS models.
For this reason, a single exponent ($y$) fully characterizes the UD-CD crossover.

% changes
If one is interested in comparison with experimental data, two other quantities are
more useful. The first one is the local roughness $w\left( r,t\right)$, which is averaged over
windows of size $r$ gliding through the surface. The second one, which will be the focus of most
of our calculations, is the HHCF of columns at distance $r$:
\begin{equation}
G\left( r,t\right) \equiv \langle {\left[ h\left( \vec{r_0}+\vec{r},t\right) -
h\left( \vec{r_0},t\right) \right]}^2 \rangle , r\equiv |\vec{r}| .
\label{defcorr}
\end{equation}
Here, the configurational averages (angular brackets) are taken over different initial positions
$\vec{r_0}$, different orientations of $\vec{r}$ and different deposits.
The local roughness and the HHCF have the same scaling properties.

For the competitive models analyzed here, the HHCF in the regime of thin
film growth ($t_c\ll t\ll t_\times$) scales as
\begin{equation}
\sqrt{G\left( r,t\right)} \approx A r^{\alpha_C} g{\left( \frac{t}{Br^{z_C}}\right)} .
\label{scalingcorrcomp}
\end{equation}
The constants $A$ and $B$ scale as the corresponding ones for the global roughness
(Eqs. \ref{defdelta} and \ref{defy}) and $g$ is a scaling function. An alternative form
that helps connection to anomalous scaling relations \cite{lopez} is
\begin{equation}
\sqrt{G\left( r,t\right)} \approx D t^{\beta_C}
G{\left( \frac{r}{{\left(t/B\right)}^{1/z_C}}\right)} ,
\label{scalingcorrcomp1}
\end{equation}
where
\begin{equation}
D \sim p^{-y\left( 1/2-\beta_C\right)} .
\label{scalingD}
\end{equation}

%small changes
For small distances $r$ ($r\ll {\left(t/B\right)}^{1/z_C}$), the HHCF is hereafter called
local HHCF and scales as
\begin{equation}
\sqrt{G_{loc}} \sim p^{-y/2}r^{\alpha_C} .
\label{Gloccomp}
\end{equation}
This quantity is time-independent in systems with normal FV scaling, but time-dependent
with anomalous scaling.
For large distances, the HHCF is hereafter called saturation HHCF and scales as
\begin{equation}
\sqrt{G_{sat}} \sim p^{-y\left( 1/2-\beta_C\right)}t^{\beta_C} .
\label{Gsatcomp}
\end{equation}
Note that $\beta_C<1/2$ for any type of CD with normal scaling, thus the exponent
$-y\left( 1/2-\beta_C\right)$ is always negative.

\section{Competitive models with decreasing rates of correlated components}
\label{competitivedecreasing}

We consider competitive models whose correlated component has time decreasing probability as
\begin{equation}
p = {\left( \frac{t}{\tau}+1\right)}^{-\Delta} ,
\label{ptime}
\end{equation}
where $\Delta$ is positive. Eq. (\ref{ptime}) is properly defined for lattice models,
so that at $t=0$ all deposition events are correlated ($p=1$). For long times
($t\gg \tau$), we have
\begin{equation}
p\sim t^{-\Delta} .
\label{ptimescaling}
\end{equation}

Eq. (\ref{ptimescaling}) can be substituted in Eqs. (\ref{Gloccomp}) and (\ref{Gsatcomp})
to give AS for the HHCF, as follows.

% small changes
The local HHCF scales with time and distance as
\begin{equation}
\sqrt{G_{loc}} \sim t^{y\Delta /2}r^{\alpha_C} .
\label{Glocptime}
\end{equation}
This quantity has the same scaling of the local roughness in small box sizes.
Thus, the local roughness exponent $\alpha_{loc}$ is the same of the correlated component,
$\alpha_C$, and the local slope exponent is
\begin{equation}
\kappa = y\Delta /2 .
\label{kappaptime}
\end{equation}
This exponent represents the degree of anomaly of the system ($\kappa =0$ for normal growth).
It depends not only on the exponent of the time-decreasing rate
but also on the aggregation conditions (BD-like or SOS) through the exponent $y$. However, it
does not depend on the universal exponents characterizing the class of the CD.

In theoretical works, $\kappa$ is sometimes called the local anomalous growth exponent
$\beta^*$ \cite{lopezPRE1996,auger}.
In experimental works, it is frequently called $\beta_{loc}$ (local growth exponent) \cite{huo},
but some authors give a different definition for $\beta_{loc}$ \cite{auger}. For these reasons,
we keep the notation close to that of Ref. \protect\cite{lopez}, consistently
with Eqs. (\ref{Glocptime}) and (\ref{kappaptime}).

% small changes
The saturation HHCF scales as
\begin{equation}
\sqrt{G_{sat}} \sim t^{\beta} \qquad , \qquad \beta = \beta_C + \Delta y\left( 1/2-\beta_C\right) .
\label{Gsatptime}
\end{equation}
This quantity has the same scaling of the local roughness in large box sizes.
Here, the global growth exponent $\beta$ is larger than the growth exponent of the CD.
In some works, that exponent is denoted as the sum of $\beta_{loc}$ and $\beta^*$ \cite{auger}.

The crossover between growth and saturation regimes of the HHCF occurs when
$r\sim {\left(t_\times/B\right)}^{1/z_C}$ (Eq. \ref{scalingcorrcomp1}). This gives the
characteristic time $t_\times$ as
\begin{equation}
t_\times \sim r^{z} \qquad ,\qquad z=\frac{z_C}{1-y\Delta} .
\label{zptime}
\end{equation}
This shows that the anomalous dynamical exponent $z$ is also larger than that of the CD,
reflecting the decreasing rate of propagation of correlations.
The expected relation
\begin{equation}
\kappa +\alpha_{loc}/z = \beta
\label{scalingrelationAS}
\end{equation}
is observed in all cases.

Due to the continuously decreasing role of the correlated dynamics, the global roughness $W$
(Eq. \ref{defw}) does not saturate, but continuously increase (i. e. there is not steady state).
This can be shown by direct
substitution of the time-dependent form of $p$ (Eq. \ref{ptime}) in the FV relation (\ref{fv}),
with the predicted forms of the scaling function and of the $p$-dependent amplitudes.

% section split in two - this is the new one
\section{Numerical results}
\label{simulations}

Three applications illustrate the remarkable effect of time-decreasing $p$, even in cases where $p$
is still of order $1$ after deposition of many layers. In all simulations, two-dimensional substrates
($d=2$) are considered.

First, we consider RDSR (also known as Family model - Fig. 1) \cite{family} as the correlated component.
This model is in the EW class, thus $\beta_C=0$ and $\alpha_C=0$,
both corresponding to logarithmic increase of the roughness in time and size, and $z_C=2$ (diffusive
dynamics) \cite{ew}. Since it is an SOS model, we have $y=2$, thus Eqs. (\ref{Gsatptime})
and (\ref{zptime}) give $\beta =\Delta$ and $z=2/\left( 1-2\Delta\right)$.

\begin{figure}
\includegraphics[width=7cm]{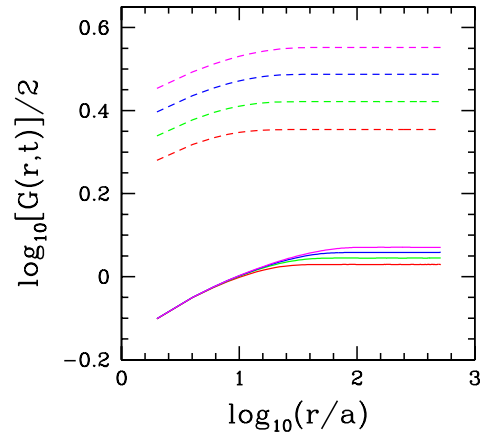}
\caption{(Color online) HHCF for the RDSR (solid curves) and for the
UD-RDSR model with $\Delta =1/6$ (dashed curves). From bottom to top, deposition times
are $t/\tau=100$ (red), $t/\tau=200$ (green), $t/\tau=400$ (blue), and $t/\tau=800$ (magenta).}
\label{fig2}
\end{figure}

Simulations of the UD-RDSR model were performed with $\Delta = 1/6$ and $\Delta =1/4$ in substrates of
size $L=1024a$ up to $t/\tau= {10}^3$. In Fig. 2, we show the
HHCF for $\Delta =1/6$ and for the pure Family model ($p=1$, $\Delta =0$) at several times.
The split of the curves for small box size $r$ is a clear signature of
anomalous scaling for $\Delta =1/6$.

\begin{figure}
\includegraphics[width=7cm]{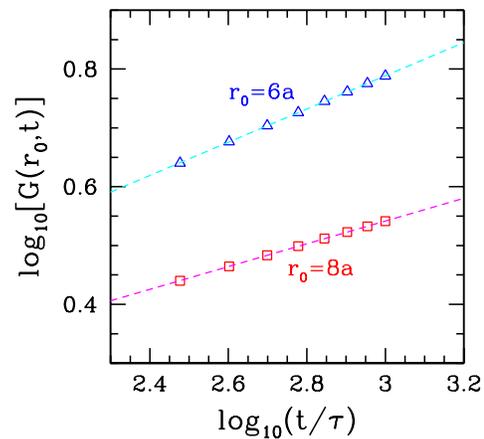}
\caption{(Color online) Local HHCF for the UD-RDSR model:
$r_0=8a$ for $\Delta = 1/6$ and $r_0=6a$ for $\Delta = 1/4$. Dashed lines are
least squares fits of the data with slopes $\kappa \approx 0.19$ and
and $\kappa \approx 0.28$, respectively.}
\label{fig3}
\end{figure}

Fig. 3 shows the time evolution of the local correlation function (fixed $r=r_0$)
for two values of $\Delta$: $r_0=8a$ for $\Delta = 1/6$ and $r_0=6a$ for $\Delta = 1/4$.
The power-law increase is clear in both cases and the linear fits give estimates of the exponent $\kappa$
slightly larger than the prediction of Eq. (\ref{kappaptime}) for $y=2$. However,
exponents calculated from local roughness or correlation function in short-time
simulations of lattice models frequently deviate from the expected values of the model class \cite{localwidth},
thus the deviation in Fig. 3 is not unexpected.

\begin{figure}
\includegraphics[width=7cm]{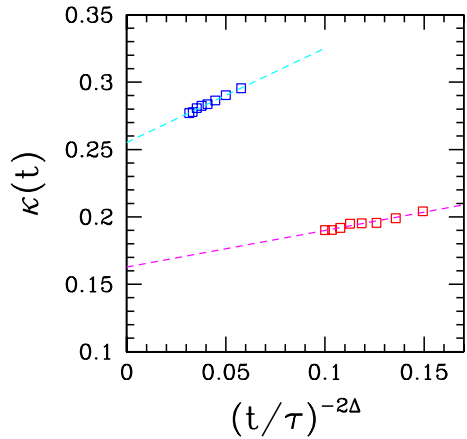}
\caption{(Color online) Effective local slope exponents, with $t_0=200\tau$,
for the UD-RDSR model: upper symbols $r_0=6a$ and $\Delta = 1/4$, lower
symbols for $r_0=8a$ and $\Delta = 1/6$. Dashed lines are least squares fits of the data.}
\label{fig4}
\end{figure}

% new paragraph, new figure 4, with effective exponents for kappa
Very accurate estimates of $\kappa$ are obtained from the effective exponents
\begin{equation}
\kappa {\left( t\right)} \equiv { \frac{\ln{\left[G_{loc}\left( r_0,t\right)
/G_{loc}\left( r_0,t-t_0\right)\right]}}{\ln{\left[ t/\left( t-t_0\right)\right]}} } ,
\label{defalphaL}
\end{equation}
which tend to $\kappa$ as $t\to\infty$, with fixed $t_0$. We assume that the leading correction
to scaling of $G_{loc}$ (not of its square root) is a constant term, in analogy to the intrinsic width
in the context of roughness scaling (see e. g. Ref. \protect\cite{kpz2d} and references therein).
Using Eqs. (\ref{Glocptime}) and (\ref{kappaptime}) with $y=2$, this assumption gives
$\kappa {\left( t\right)}\to \Delta$ with a correction term proportional to $t^{-2\Delta}$. 
In Fig. 4, we plot $\kappa {\left( t\right)}$ as a function of $t^{-2\Delta}$ using the same data
of Fig. 3 with  $t_0=200\tau$. The linear fits give asymptotic estimates $\kappa =0.255$ for
$\Delta=0.25$ and $\kappa =0.163$ for $\Delta =1/6=0.1666\dots$, in excellent agreement with
the theoretical prediction $\kappa = \Delta$.

Now we consider BD as the correlated component. In BD, the incident particle attaches to
the site where it has a nearest neighbor occupied site (at the sides or below it) \cite{barabasi,vold},
creating a porous deposit due to lateral aggregation.
The model is in the KPZ class \cite{kpz}, where  $\alpha_C\approx 0.385$, $z_C\approx 1.615$,
and $\beta_C\approx 0.24$ \cite{marinari,colaiori,kpz2d}.
Using $y=1$ for the competitive model, Eqs. (\ref{Gsatptime}) and (\ref{zptime}) give
$\beta\approx 0.24+0.26\Delta$ and $z\approx 1.615/\left( 1-\Delta\right)$. 

\begin{figure}
\includegraphics[width=7cm]{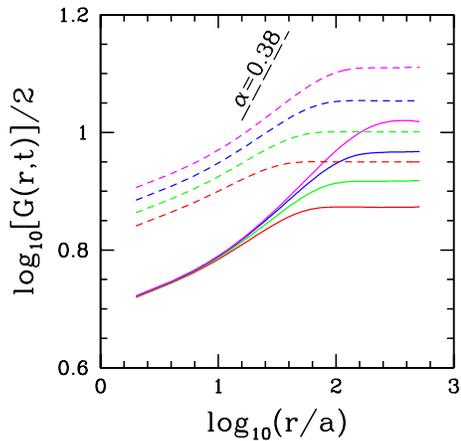}
\caption{(Color online) HHCF for BD (solid curves) and for the 
UD-BD model with $\Delta =1/6$ (dashed curves). From bottom to top, deposition times
are $t/\tau=100$ (red), $t/\tau=200$ (green), $t/\tau=400$ (blue), and $t/\tau=800$ (magenta).
The long-dashed line has the expected slope of the KPZ class.}
\label{fig5}
\end{figure}

Simulations of the UD-BD model were performed in two-dimensional substrates of
size $L=1024a$ up to $t/\tau= {10}^3$, again for $\Delta = 1/6$ and $\Delta =1/4$.
Fig. 5 shows the HHCF for $\Delta =1/6$ and for the pure BD model
($p=1$, $\Delta =0$) at several times. Comparison with Fig. 2 shows a smaller split
of the curves for small $r$, which is expected for the smaller exponent $y$ (Eq. \ref{Glocptime}).

An interesting effect shown in Fig. 5 is the small slope of the curves in the scaling region,
which gives local roughness exponents much smaller than the expected
$\alpha_{loc}\approx \alpha_C\approx 0.38$.
The apparent roughness exponent is even smaller in the competitive model.
However, large discrepancies between exponents from
simulations of BD-like models and KPZ values are frequent \cite{sigma}, thus the deviations in
Fig. 5 are not a particular feature of models with AS.

A third model used to illustrate our results has the restricted SOS (RSOS) model
\cite{kk} as the correlated component. The condition for aggregation of a particle at the
column of incidence is that the height difference between all neighboring columns cannot
exceed one lattice unit, otherwise the aggregation attempt is rejected.
This means that the aggregation is accepted only if the height of the column of incidence
is smaller than or equal to the heights of all neighbors (this rule is used in the extension
to the competitive model). The RSOS model also belongs to the KPZ class and usually
shows small scaling corrections.
Using $y=2$ for the competitive model and estimates of KPZ exponents  \cite{marinari,colaiori,kpz2d},
Eqs. (\ref{Gsatptime}) and (\ref{zptime}) give
$\beta\approx 0.24+0.52\Delta$ and $z\approx 1.615/\left( 1-2\Delta\right)$. 

\begin{figure}
\includegraphics[width=7cm]{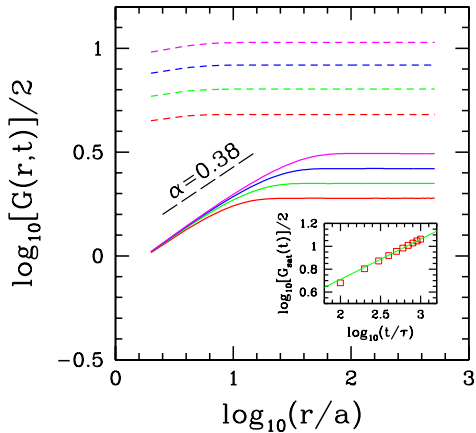}
\caption{(Color online) HHCF for the RSOS model (solid curves) and for the
UD-RSOS model with $\Delta =1/6$ (dashed curves). From bottom to top, deposition times
are $t/\tau=100$ (red), $t/\tau=200$ (green), $t/\tau=400$ (blue), and $t/\tau=800$ (magenta).
Inset: saturation HHCF as a function of time,  with a linear fit of the last
five points (solid curve) with slope $\beta\approx 0.35$.}
\label{fig6}
\end{figure}

Simulations of the UD-RSOS model are performed
in two-dimensional substrates under the same conditions of the previous ones.
Fig. 6 shows the HHCF for $\Delta =1/6$ at times between $t/\tau =100$ and $t/\tau =800$.
Again, AS is clearly shown by the split of the curves for small $r$, in contrast to the case
of time-independent rates.

The inset of Fig. 6 shows the time scaling of $G_{sat}$ for the competitive model, which gives
$\beta\approx 0.35$ for the longer times. This is in good agreement with the theoretical prediction
$\beta\approx 0.33$. Again, extrapolation of effective exponents (not shown) provide
estimates much closer to the theoretical prediction.
 
A strinking feature of this model is the small slope of the
$\log{G}\times \log{r}$ curves for small $r$, caused by the large height fluctuation
produced by UD, in comparison with the typically small roughness of the RSOS model. This feature
masks the asymptotic scaling for small $r$ in the model with AS.

\section{EW and KPZ equations}
\label{equations}

% small changes
For competitive models such as UD-RDSR, the general form of the associated EW equation can be derived,
with coefficients related to $p$ and scaling exponents related to $\Delta$. Consider, for instance,
the case in two-dimensional substrates ($d=2$). The surface tension term of the EW equation \cite{ew}
is expected to decrease in time, thus we propose
\begin{equation}
\frac{\partial h}{\partial t} = \nu_0 {\left( \frac{t}{t_0}\right)}^{-\Omega}
{\nabla}^2 h + \eta (\vec{x},t) ,
\label{ewcomp}
\end{equation}
where $h$ is the height at the position $\vec{x}$ in the
$d$-dimensional substrate at time $t$, $\nu_0$ and $t_0$ are constants, and $\eta$ is a Gaussian
noise~\cite{barabasi,kpz} with zero mean and co-variance $\langle
\eta\left(\vec{x},t\right) \eta (\vec{x'},t')\rangle = D\delta^d
(\vec{x}-\vec{x'} ) \delta\left( t-t'\right)$.

% changes
A complete representation of the lattice model
would require calculation of the constants $\nu_0$ and $t_0$, as well as the coefficients of 
higher-order derivatives not shown in Eq. (\ref{ewcomp}). However, that form is
sufficient for calculation of scaling exponents. Indeed,
scaling arguments following the same lines of Ref. \protect\cite{barabasi} lead to
$z=2/\left( 1-\Omega\right)$ and global roughness exponent $\alpha_{GL} = \Omega/\left( 1-\Omega\right)$
($\alpha_{GL}=\alpha_{loc}+z\kappa$ \cite{lopez}).
Since $z_C=2$ and $\alpha_C =0$ for the normal EW equation (with coefficients not changing in time)
and $y=2$ (SOS model), comparison with Eq. (\ref{zptime}) leads to $\Omega = 2\Delta$.
Equivalently, the full surface tension coefficient is $\nu \sim p^2$
(as predicted in Ref. \protect\cite{lam} in $1+1$ dimensions).
Similar result is obtained for lattice models in the EW class in any other substrate dimension.

% small changes 
On the other hand, the KPZ equation \cite{kpz} corresponding to
such competitive models can be constructed only in $d=1$:
\begin{equation}
\frac{\partial h}{\partial t} = \nu\left( t\right){\nabla}^2 h + \frac{\lambda\left( t\right)}{2}
{\left( \nabla h\right) }^2 + \eta (\vec{x},t) .
\label{kpz}
\end{equation}
The roughness amplitudes of Eqs. (\ref{fv}) and (\ref{scalingttimes}) scale
with the coefficients of the linear and nonlinear terms as $A\sim \nu^{-1/2}$ and $B\sim |\lambda|\nu^{-1/2}$
\cite{af}. Using Eqs. (\ref{defdelta}), (\ref{defy}), and (\ref{ptime}),
for BD-like models, we have $\nu\sim t^{-\Delta}$ and $\lambda\sim t^{-3\Delta /2}$.
However, for SOS models ($y=2$) in the KPZ class, such as UD-RSOS, we have
$\nu\sim t^{-2\Delta}$ and $\lambda\sim t^{-3\Delta}$. Again, a complete representation of the
lattice model by a stochastic equation would require calculation of coefficients of the terms in
Eq. (\ref{kpz}) (not only their scaling properties) and of higher-order terms not shown there.

\section{Discussion}
\label{discussion}

The results for $\Delta = 1/6$ show that anomalous scaling can be observed in competitive growth
models even with very slow changes of the rates and far from the crossover region $p\ll 1$.
At $t=800\tau$ (Figs. 2, 4 and 5), we have $p\approx 0.33$, i. e. there is a significant contribution of the
correlated process.
For comparison, numerical work on competitive models with fixed $p$ usually focus the regime $p\leq 0.2$
\cite{albano1,albano2,rdcor}.
Thus, the time-dependence of the competitive dynamics is the main ingredient for the AS,
independently of approaching the crossover region.

Although the competitive models presented here are drastic simplifications of a real thin film growth
process, their assumptions are not far from experimentally feasible conditions.
If one assumes that the correlated aggregation is an activated process, then the decrease of the
probability $p$ may be a consequence of decreasing the temperature during
the film growth. In the case of an EW or KPZ model, that assumption suggests an Ahrrenius form
for the surface tension coefficient as
$\nu\sim \exp{\left( -E/k_BT\right)}$, which gives $p\sim \nu^{1/y}\sim \exp{\left( -E/yk_BT\right)}$
($y=1$ or $2$). If other processes produce correlations (e. g. surface diffusion), then the coefficients
of the corresponding growth equation will also have the Ahrrenius form.
Now, assuming that $E=0.2 eV$, a temperature decrease from $330 K$ to $300 K$ gives a ratio of
probabilities $\approx 0.7$ for $y=2$. For comparison, in Figs. 2, 4  and 5 ($\Delta = 1/6$),
the probability $p$ decreases by a factor $0.71$ from $t=100\tau$ to $t=800\tau$. 

Experimental measurement of anomalous exponents $\kappa$, $\beta$ and $z$ allows testing the
hypothesis of a competition of uncorrelated and correlated deposition. From Eqs. (\ref{kappaptime}),
(\ref{Gsatptime}), and (\ref{zptime}), their values provide estimates of the scaling exponents
of the correlated component as
\begin{equation}
\Delta = 2\kappa /y \qquad ,\qquad \beta_C = \frac{\beta -\kappa}{1-2\kappa} \qquad ,\qquad
z_C = z{\left( 1-2\kappa\right)} .
\label{exprelations}
\end{equation}
The estimates of $\beta_C$ and $z_C$ can be compared with existing theories of interface growth. Note
that they are measured independently of the exponent $y$, thus it is not necessary a priori knowledge
of the aggregation mechanism.

Applications of this model are not expected when $\kappa$ is very close to $1/2$.
For instance, if $\kappa =0.4$, we obtain $z=5z_C$ and $\beta -\kappa=5\beta_C$, i. e.
the anomalous exponents are much larger than those of theories of normal scaling (the difference
$\beta -\kappa$ is frequently referred as $\beta$ in experimental works). For this reason, the
AS with large $\kappa$ observed in metal electrodeposition by some authors has to be explained by other
approaches \cite{huo,saitou}. Other systems showing AS with large values of $\kappa$
\cite{hasan,auger,fu}, as well as experiments on metal dissolution \cite{cordobaPRL2009,cordobaPRE2008},
will also give unreliable estimates of $\beta_C$ and/or $z_C$ if our competitive model is applied.

On the other hand, the work on $Cu$ electrodeposition by Lafouresse et al \cite{lafouresse}
suggests an application. At the early stages of growth and for low potential values,
the exponents $\kappa=0.07$, $\beta -\kappa=0.21$ and $\alpha_{loc}\sim 1$ are obtained. Using our
model, we find $\beta_C=0.24$ and $z=4.1$. These values are very close to the exponents
$\beta_C=0.25$ and $z_C=4$ of the linear fourth-order stochastic growth equation
\begin{equation}
\frac{\partial h}{\partial t} = \nu_4{\nabla}^4 h + \eta (\vec{x},t) 
\label{mh}
\end{equation}
(Mullins-Herring class \cite{barabasi,mh}), which represents growth dominated by surface diffusion.
Consequently, a possible interpretation of the experiment under those conditions
is that the role of the surface diffusion
is continuously decreasing in time due to a competition with an uncorrelated random growth.

The results for lattice models shown in Figs. 4 and 5 also suggest that deviations in
the small $r$ scaling are frequent. Consequently, when experimental data is analyzed and
compared to our model, one should consider that a disagreement in the values of $\alpha$
may not represent a failure of the model. Instead, this is an expected feature at short
times / small thicknesses \cite{localwidth}.

\section{Conclusion}

We studied lattice models with competition of correlated and uncorrelated mechanisms for
aggregation of deposited particles and where the rate of the correlated component decrease in time as a
power-law. The Family-Vicsek scaling relation derived for the case of constant rates is
extended by direct substitution of the time-dependent rate, giving anomalous scaling relations
that are supported by simulation data. The anomalous exponents are related to the exponents of
the correlated growth dynamics and to the exponent of the time-decreasing rate.
The EW and KPZ equations corresponding to some models are discussed.

Even with a slow decay of the correlated component, which may correspond to small
temperature changes in systems with activated dynamics, a remarkable anomaly is found.
Indeed, application of the model is expected only for systems with small anomaly exponent
$\kappa$ ($\beta^*$ or $\beta_{loc}$ in some works). This excludes most electrodeposition works,
but a case of thin $Cu$ films electrodeposited at low potential values shows a possible
application of growth dominated by surface diffusion.
Since our scaling relations provide exponents of the correlated component depending only on
anomalous exponents that can be measured experimentally (independenly of the exponents
of the time-decreasing rate and of the aggregation mechanism), the test in other real systems
is simple and may eventually help to understand their basic physical mechanisms.

\acknowledgements

This work was partially supported by CNPq and FAPERJ (Brazilian agencies).

%~~~~~~~~~~~~~~~~~~~~~~~~~~~~~~~~~~~~~~~~~~~~~~~~~~~~~~~~~~~~~~~~~~~~~~~~~~~
%~~~~~~~~~~~~~~~~~~~  REFERENCES  ~~~~~~~~~~~~~~~~~~~~~~~~~~~~~~~~~~~~~~~~~~
%~~~~~~~~~~~~~~~~~~~~~~~~~~~~~~~~~~~~~~~~~~~~~~~~~~~~~~~~~~~~~~~~~~~~~~~~~~~

\end{document}